\documentclass[conference]{IEEEtran}
\IEEEoverridecommandlockouts

\usepackage{cite}
\usepackage{amsmath,amssymb,amsfonts}
\usepackage{bm}
\usepackage{graphicx}
\usepackage{textcomp}
\usepackage{xcolor}
\usepackage{booktabs}
\usepackage{multirow}

\def\BibTeX{{\rm B\kern-.05em{\sc i\kern-.025em b}\kern-.08em
    T\kern-.1667em\lower.7ex\hbox{E}\kern-.125emX}}

\begin{document}


\setlength{\abovedisplayskip}{3pt}
\setlength{\belowdisplayskip}{3pt}
\setlength{\abovedisplayshortskip}{2pt}
\setlength{\belowdisplayshortskip}{2pt}
\setlength{\textfloatsep}{6pt}
\setlength{\floatsep}{5pt}
\setlength{\intextsep}{5pt}


\title{
Green Learning-Based Signal-Level Detection of RIS Spoofing Attacks in Vehicular ISAC Systems
\vspace{-0.1in}

\thanks{This work was supported in part by the National Science and Technology Council (NSTC) of Taiwan under Grant 113-2926-I-001-502-G, 114-2221-E-A49-185-MY3 and 115-2221-E-A49-073-MY3.
Prof. Walid Saad was supported by the U.S. National Science Foundation (NSF) under Grant CNS-2114267.}
}

\author{
\IEEEauthorblockN{
Po-Heng Chou$^{1,2,3}$,
Tingyu Shui$^{1}$,
Pin-Qi Fu$^{4}$,
Lien-Sheng Chang$^{5}$,
Walid Saad$^{1}$,
and Li-Chun Wang$^{3,4}$
}
\IEEEauthorblockA{
$^{1}$Bradley Department of Electrical and Computer Engineering (ECE),
Virginia Tech (VT), Alexandria, VA 22305, USA\\
$^{2}$Research Center for Information Technology Innovation (CITI),
Academia Sinica (AS), Taipei 11529, Taiwan\\
$^{3}$Department of Electronics and Electrical Engineering and $^{4}$Institute of Communications Engineering,\\ National Yang Ming Chiao Tung University (NYCU), Hsinchu 30010, Taiwan\\
$^{5}$Institute of Communication Engineering (ICE), National Sun Yat-sen University (NSYSU), Kaohsiung 80424, Taiwan\\
E-mails: d00942015@ntu.edu.tw,
tygrady@vt.edu,
amaza.ee12@nycu.edu.tw,
willy22559@gmail.com,\\
walids@vt.edu,
wang@nycu.edu.tw
\vspace{-0.25in}
}
}

\maketitle


\begin{abstract}
Integrated sensing and communication (ISAC) enables wireless
infrastructure to jointly support communication and environmental
sensing, but also introduces new physical-layer security threats. A
malicious reconfigurable intelligent surface (RIS) can manipulate its
time-varying reflection phases to generate spoofing echoes and distort
the sensing results of a roadside unit (RSU). Existing trajectory-level
detectors rely on reconstructed sensing parameters and vehicle
trajectories, making their performance dependent on upstream estimation
accuracy. In this paper, we propose a signal-level Green Learning (GL)
framework that directly detects RIS spoofing from matched-filtered RSU
echoes. The proposed method combines two-stage Subspace Approximation
with Adjusted Bias (SAAB), Discriminant Feature Test (DFT), and
eXtreme Gradient Boosting (XGBoost) for spatial-temporal
representation learning, feature selection, and classification,
respectively. Simulation results show 99.335\% detection accuracy for
a 32-element malicious RIS at $-80$ dBm receiver noise power.
GL remains competitive with Long Short-Term Memory (LSTM) and
bidirectional LSTM (BiLSTM) baselines while requiring only 0.1119
million floating-point operations (FLOPs), and also generalizes to
previously unseen RIS sizes.
\end{abstract}

\begin{IEEEkeywords}
Green Learning, integrated sensing and communication,
reconfigurable intelligent surface, sensing security,
spoofing detection, vehicular networks.
\end{IEEEkeywords}


\section{Introduction}

Integrated sensing and communication (ISAC) is expected to become a
key capability of sixth-generation (6G) wireless networks by enabling
a common radio infrastructure to support both communication and
environmental sensing. The International Telecommunication Union Radiocommunication Sector
(ITU-R) has identified integrated sensing and communication as an
important capability toward International Mobile Telecommunications
(IMT)-2030~\cite{itur2160}. Recent surveys have further highlighted the
architectural, technological, standardization, and experimental
development of ISAC toward practical next-generation wireless
systems~\cite{isacsurvey2025}.

Vehicular networks are a representative application of ISAC because
communication and sensing information can be obtained from the same
roadside infrastructure. A roadside unit (RSU) can communicate with a
vehicular user (VU) while simultaneously estimating its position and
velocity from reflected signals~\cite{du2025vehicular}. Such sensing information can support beam tracking, localization,
cooperative driving, and collision avoidance, making sensing
reliability critical to downstream communication and control.

Millimeter-wave (mmWave) operation provides large bandwidth and high
sensing resolution, but also suffers from severe propagation loss and
blockage. Reconfigurable intelligent surface (RIS) technology can
mitigate these limitations by creating controllable propagation paths
through programmable passive reflecting elements
~\cite{du2025vehicular}. However, the same programmability also
creates a new physical-layer attack surface.

A malicious RIS can deliberately manipulate its time-varying
reflection phases to alter the sensing echo received at the RSU.
Unlike conventional jamming, which primarily degrades the received
signal-to-noise ratio (SNR), RIS-based spoofing can generate
structured artificial echoes that resemble physically plausible
sensing responses. Consequently, the sensing receiver may infer
incorrect delay, Doppler, angle, or trajectory information even when
the received signal remains apparently consistent with normal
operation.

Existing physical-layer security studies for ISAC have mainly focused
on protecting communication confidentiality or mitigating malicious
interference. Secure radar-communication transmission against
malicious sensing targets and eavesdroppers has been investigated
through waveform, beamforming, and interference design~\cite{su2021secure,su2022secure}. Sensing information has also been exploited to improve secure
communication by acquiring sensing-assisted information about
potential eavesdroppers~\cite{cao2025sensing}. More recently,
research has begun to directly address attacks against the sensing
functionality. Orthogonal frequency-division multiplexing (OFDM)
replay attacks can generate false range-Doppler targets~\cite{chrysanidis2024replay}, while intelligent surfaces can be used
to suppress or manipulate unauthorized sensing measurements~\cite{xu2025stealth}. RIS-assisted ISAC has also been investigated
from a broader cyber-physical security perspective~\cite{hassouna2026riscs}.

Recently, our work~\cite{shui2025safety} investigated
RIS-based sensing spoofing in vehicular ISAC systems. The authors
showed that a malicious RIS can manipulate its time-varying phase
shifts to distort Doppler and angle-of-departure (AoD) estimation at
the RSU. They derived the feasible spoofing-frequency region and
analyzed the resulting sensing errors. This physical-layer attack
model was subsequently extended in~\cite{shui2025spoofing}, where
spoofed-trajectory generation was formulated as a Markov decision
process (MDP) and a signal temporal logic (STL)-based neuro-symbolic
detector was developed for trajectory-level attack detection.

This paper instead investigates RIS spoofing detection directly at the
received-signal level. Building on the physical attack model and
attack-generation framework in~\cite{shui2025safety,shui2025spoofing}, the proposed detector
operates on matched-filtered RSU echoes before complete
sensing-parameter estimation and trajectory reconstruction. This formulation avoids requiring reconstructed sensing parameters
and trajectories as detector inputs and enables spoofing to be identified earlier in
the sensing processing chain.

Recurrent neural networks provide a natural baseline for signal-level
sequence classification. In particular, Long Short-Term Memory (LSTM)
networks can model temporal dependencies across sensing
observations~\cite{hochreiter1997lstm}. However, recurrent matrix
operations and backpropagation-based training introduce considerable
computational and energy costs, which are undesirable for continuously
operating RSUs and resource-constrained edge sensing platforms.

Green Learning (GL) provides a lightweight alternative by separating
representation learning, feature selection, and decision learning.
Subspace Approximation with Adjusted Bias (SAAB) provides a
feedforward, data-driven representation without iterative
backpropagation~\cite{kuo2019interpretable}. Discriminant Feature Test
(DFT) then ranks candidate features according to their class
discrimination capability~\cite{yang2022supervised}. Combined with
tree-based decision learning, this modular design can reduce
computational complexity while preserving discriminative signal
information~\cite{kuo2022green}.

The main contributions of this work are summarized as follows:

\begin{itemize}
\setlength{\itemsep}{0pt}
\setlength{\parsep}{0pt}
\setlength{\topsep}{2pt}

\item Building on the physically grounded RIS spoofing models in~\cite{shui2025safety,shui2025spoofing}, we formulate the defense
problem at the received-signal level. The detector operates directly
on matched-filtered RSU echoes before complete sensing-parameter
estimation and trajectory reconstruction.

\item We develop a three-stage GL detector consisting of two-stage
SAAB-based spatial-temporal representation learning, DFT-based
supervised feature selection, and eXtreme Gradient Boosting (XGBoost)
classification~\cite{chen2016xgboost}.

\item We evaluate the proposed detector under different receiver noise
powers and malicious RIS sizes. Cross-condition experiments further
quantify generalization when a detector trained at one RIS size is
directly evaluated at previously unseen RIS sizes.

\item We quantify both arithmetic complexity and measured energy
consumption. The proposed GL detector requires only 0.1119 million
floating-point operations (FLOPs), substantially below the considered
LSTM~\cite{hochreiter1997lstm} and bidirectional LSTM (BiLSTM)~\cite{schuster1997bidirectional}  baselines while maintaining
competitive detection accuracy.

\end{itemize}


\section{System Model and RIS Spoofing Attack}

\subsection{Vehicular ISAC Signal Model}

\begin{figure}[t]
\centering
\includegraphics[width=0.85\columnwidth]
{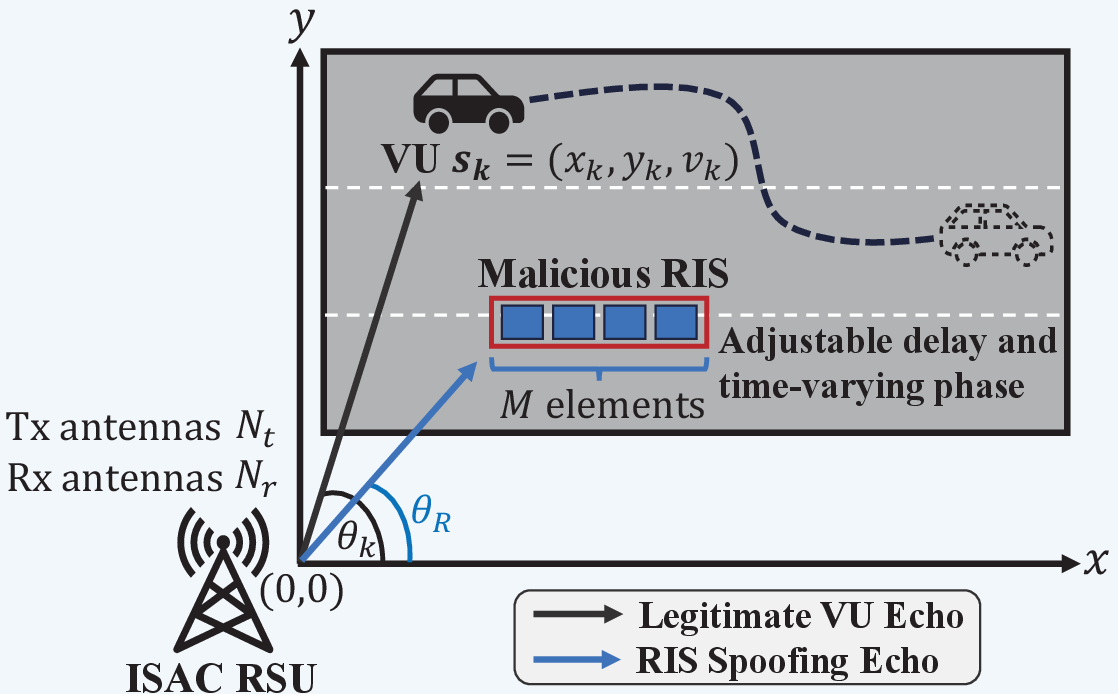}
\caption{Vehicular ISAC system in the presence of a malicious RIS.}
\label{fig_system}
\end{figure}

Fig.~\ref{fig_system} illustrates the considered vehicular ISAC
system in the presence of a malicious RIS. The system consists of a
full-duplex mmWave RSU, a single-antenna VU, and a malicious RIS
~\cite{shui2025safety,shui2025spoofing}. The RSU employs a uniform
linear array (ULA) with $N_t$ transmit antennas and $N_r$ receive
antennas, whereas the RIS contains $M$ reflecting elements.

Lowercase bold symbols denote vectors and uppercase bold symbols denote
matrices. The sets of complex and real numbers are denoted by
$\mathbb{C}$ and $\mathbb{R}$, respectively. The operators
$(\cdot)^T$, $(\cdot)^H$, $(\cdot)^*$, and
$\operatorname{diag}(\cdot)$ denote transpose, Hermitian transpose,
complex conjugation, and the diagonal-matrix operator, respectively.

The VU motion is represented over $K$ sensing slots, each with
duration $T$. At slot $k\in\{1,\ldots,K\}$, its state is $\bm{s}_k=[x_k,y_k,v_k]^T$, where $x_k$ and $y_k$ are the Cartesian
coordinates and $v_k$ is the velocity.
Let $t$ denote continuous time within a sensing slot.
The waveform transmitted during slot $k$ is
$q_k(t)\in\mathbb{C}$ with $q_k(t)q_k^*(t)=1$, and let
$\bm{w}_k\in\mathbb{C}^{N_t\times1}$ denote the RSU transmit
precoding vector.

The legitimate VU echo is modeled as
\begin{equation}
\bm{y}_{V,k}(t)
=
\sqrt{P}\gamma_B\beta_{V,k}e^{j2\pi\mu_k t}
\bm{b}_B(\theta_k)
\bm{a}_B^H(\theta_k)
\bm{w}_kq_k(t-\tau_k),
\label{eq_vu}
\end{equation}
where $P$ is the RSU transmit power, $j=\sqrt{-1}$,
$\gamma_B=\sqrt{N_tN_r}$ is the RSU array-gain factor,
$\beta_{V,k}$ is the complex VU reflection coefficient,
$\mu_k$ is the Doppler shift, $\theta_k$ is the VU direction, and
$\tau_k$ is the round-trip delay.

\begin{figure*}[t]
\centering
\includegraphics[width=\textwidth]
{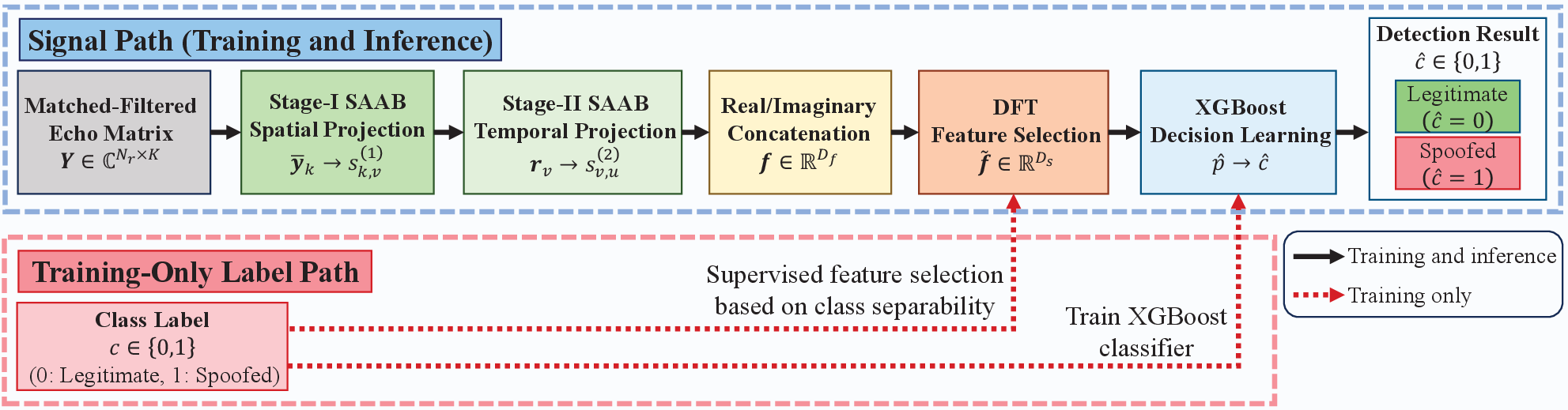}
\caption{Proposed SAAB--DFT--XGBoost signal-level detection framework.}
\label{fig_gl}
\vspace{-0.2in}
\end{figure*}

For a generic direction $\theta$ and half-wavelength spacing, the normalized RSU steering vectors are
$\bm{a}_B(\theta)=N_t^{-1/2}
[1,e^{-j\pi\cos\theta},\ldots,e^{-j\pi(N_t-1)\cos\theta}]^T$
and
$\bm{b}_B(\theta)=N_r^{-1/2}
[1,e^{-j\pi\cos\theta},\ldots,e^{-j\pi(N_r-1)\cos\theta}]^T$.
The VU reflection coefficient is
$\beta_{V,k}=
\sqrt{\lambda^2\kappa_V/(64\pi^3d_k^4)}
e^{j4\pi d_k/\lambda}$,
where $\lambda$ is the carrier wavelength, $\kappa_V$ is the VU radar
cross section (RCS), and $d_k$ is the RSU-to-VU distance.

The malicious RIS generates the additional echo
\begin{align}
\bm{y}_{R,k}(t)
&=
\sqrt{P}\gamma_B\beta_R
\bm{b}_B(\theta_R)
\bm{a}_R^H(\theta_B)
\bm{\Phi}_k(t)
\bm{b}_R(\theta_B)
\nonumber\\
&\quad\times
\bm{a}_B^H(\theta_R)
\bm{w}_kq_k(t-\tau_R),
\label{eq_ris}
\end{align}
where $\beta_R$ is the complex RIS reflection coefficient,
$\theta_R$ is the RIS direction observed from the RSU,
$\theta_B$ is the RSU direction observed from the RIS, and
$\tau_R$ is the effective RSU--RIS--RSU round-trip delay.
The normalized RIS steering vectors
$\bm{a}_R(\theta),\bm{b}_R(\theta)\in\mathbb{C}^{M\times1}$
are defined analogously to the RSU steering vectors.

The RIS applies the time-varying reflection matrix
$\bm{\Phi}_k(t)=
\operatorname{diag}(e^{j\phi_{k,1}(t)},\ldots,e^{j\phi_{k,M}(t)})$,
where $\phi_{k,m}(t)$ is the programmable phase of element
$m\in\{1,\ldots,M\}$.
The RIS reflection coefficient is
$\beta_R=
\sqrt{\lambda^2\kappa_R/(64\pi^3d_R^4)}
e^{j4\pi d_R/\lambda}$,
where $d_R$ is the RSU-to-RIS distance. Following
~\cite{shui2025safety,shui2025spoofing},
$\kappa_R=4\pi\eta S^2/\lambda^2$, where $\eta$ is the RIS reflection
efficiency and $S$ is its physical area.

The received sensing signal is
\begin{equation}
\bm{y}_k(t)
=
\bm{y}_{V,k}(t)
+
\bm{y}_{R,k}(t)
+
\bm{z}_k(t),
\label{eq_received}
\end{equation}
where
$\bm{z}_k(t)\sim
\mathcal{CN}(\bm{0},\sigma^2\bm{I}_{N_r})$
is additive white Gaussian noise (AWGN),
$\mathcal{CN}(\cdot,\cdot)$ denotes a circularly symmetric complex
Gaussian distribution, $\sigma^2$ is the receiver noise power, and
$\bm{I}_{N_r}$ is the $N_r\times N_r$ identity matrix.

\subsection{RIS Spoofing Mechanism}

The malicious RIS aims to shift the dominant matched-filter peak in
the Doppler domain. For candidate delay $\tau$ and Doppler frequency
$\mu$, the RSU computes
\begin{equation}
C_k(\tau,\mu)
=
\left\|
\int_0^T
\bm{y}_k(t)
q_k^*(t-\tau)
e^{-j2\pi\mu t}
dt
\right\|_2^2,
\label{eq_matched}
\end{equation}
where $C_k(\tau,\mu)$ is the matched-filter output power and
$\|\cdot\|_2$ denotes the Euclidean norm. The delay and Doppler
estimates are
\begin{equation}
(\hat{\tau}_k,\hat{\mu}_k)
=
\arg\max_{\tau,\mu}
C_k(\tau,\mu).
\label{eq_estimate}
\end{equation}

Following~\cite{shui2025safety,shui2025spoofing}, an adjustable-delay
RIS is assumed such that the malicious echo overlaps the legitimate
VU echo in the delay domain, i.e., $\tau_R\simeq\tau_k$. The RIS then
manipulates the apparent Doppler through the common phase law
$\phi_{k,m}(t)=2\pi\tilde{\mu}_k t$, where $\tilde{\mu}_k$ is the
attacker-selected apparent Doppler frequency. The corresponding
Doppler offset is
$\Delta\tilde{\mu}_k=\tilde{\mu}_k-\mu_k$.

A spoofing offset is feasible when the malicious matched-filter peak
is at least as strong as the legitimate peak. Hence,
\begin{equation}
\mathcal{A}_k
=
\left\{
\Delta\tilde{\mu}_k:
C_k\!\left(
\hat{\tau}_k,
\mu_k+\Delta\tilde{\mu}_k
\right)
\geq
C_k(\hat{\tau}_k,\mu_k)
\right\},
\label{eq_attack_set}
\end{equation}
where $\mathcal{A}_k$ denotes the set of feasible Doppler offsets.

As shown in~\cite{shui2025safety,shui2025spoofing}, coherent
combination over the $M$ RIS elements yields an approximately $M^2$
spoofing-echo power gain, so increasing $M$ strengthens the malicious
echo and can enlarge $\mathcal{A}_k$.

The proposed detector operates directly on matched-filtered echoes.
After delay-Doppler compensation, the processed echo at slot $k$ is
\begin{equation}
\bar{\bm{y}}_k
=
\int_0^T
\bm{y}_k(t)
q_k^*(t-\hat{\tau}_k)
e^{-j2\pi\hat{\mu}_k t}
dt
\in\mathbb{C}^{N_r\times1}.
\label{eq_processed_echo}
\end{equation}
The observations over $K$ slots are stacked as
\begin{equation}
\bm{Y}
=
[
\bar{\bm{y}}_1,\ldots,\bar{\bm{y}}_K
]
\in\mathbb{C}^{N_r\times K}.
\label{eq_input}
\end{equation}
Let $c\in\{0,1\}$ denote the class label, where $c=0$ and $c=1$
represent legitimate and RIS-spoofed observations, respectively.
The detector outputs $\hat{c}=g(\bm{Y})$, where $g(\cdot)$ denotes
the proposed classifier.


\section{Proposed Green Learning Detector}

Fig.~\ref{fig_gl} illustrates the proposed signal-level GL detection
framework. The detector consists of three main stages: SAAB-based
representation learning, DFT-based supervised feature selection, and
XGBoost-based decision learning. Specifically, two-stage SAAB first
extracts spatial and temporal signal structure, DFT retains the most
discriminative features, and XGBoost performs the final binary
classification.

\subsection{Two-Stage SAAB Representation Learning}

The input
$\bm{Y}=[\bar{\bm{y}}_1,\ldots,\bar{\bm{y}}_K]$
contains spatial variation across the $N_r$ receive antennas and
temporal variation across the $K$ sensing slots. SAAB is therefore
applied sequentially along these two dimensions.

In the first stage, each
$\bar{\bm{y}}_k\in\mathbb{C}^{N_r\times1}$ is treated as a spatial
sample. Following~\cite{kuo2019interpretable}, the direct-current
(DC) anchor is
$\bm{a}_0^{(1)}=N_r^{-1/2}[1,\ldots,1]^T$.
After removing the DC component, eigendecomposition of the training
covariance matrix yields the alternating-current (AC) anchors
$\bm{a}_v^{(1)}$. The corresponding spatial coefficient is
\begin{equation}
s_{k,v}^{(1)}
=
\left(\bm{a}_v^{(1)}\right)^H
\bar{\bm{y}}_k,
\label{eq_saab1}
\end{equation}
where $v\in\{1,\ldots,V_1\}$ indexes the retained spatial components
and $V_1$ denotes their number. The dominant $V_1$
components are retained according to projection energy.

The second stage captures temporal variation. For retained component
$v$, define
$\bm{r}_v=
[s_{1,v}^{(1)},\ldots,s_{K,v}^{(1)}]^T
\in\mathbb{C}^{K\times1}$.
Its temporal SAAB coefficient is
\begin{equation}
s_{v,u}^{(2)}
=
\left(\bm{a}_u^{(2)}\right)^H
\bm{r}_v,
\label{eq_saab2}
\end{equation}
where $\bm{a}_u^{(2)}\in\mathbb{C}^{K\times1}$ is the $u$th temporal
anchor and where $u\in\{1,\ldots,V_2\}$ indexes the retained temporal
components and $V_2$ denotes their number.

Once the anchors are learned from the training data, SAAB inference
requires only fixed projections. Let $\bm{s}^{(2)}$ collect the
retained second-stage coefficients. Their real and imaginary parts are
concatenated into
$\bm{f}=
[\operatorname{Re}\{\bm{s}^{(2)}\}^T,
\operatorname{Im}\{\bm{s}^{(2)}\}^T]^T
\in\mathbb{R}^{D_f}$,
where $\operatorname{Re}\{\cdot\}$ and
$\operatorname{Im}\{\cdot\}$ denote the real and imaginary parts and
where $D_f=2V_1V_2$ is the resulting real-valued feature dimension.

\subsection{DFT-Based Supervised Feature Selection}

Although SAAB compresses the received echoes, the retained coefficients
are not equally useful for distinguishing legitimate and spoofed
samples. DFT therefore ranks them according to their supervised
discrimination capability~\cite{yang2022supervised}.

For feature $i\in\{1,\ldots,D_f\}$ and threshold $t$, let
$N_L^{i,t}$ and $N_R^{i,t}$ denote the numbers of training samples
on the two sides of the threshold, and let
$p_{j,L}^{i,t}$ and $p_{j,R}^{i,t}$ denote the corresponding
proportions of class $j\in\{0,1\}$. The weighted entropy is
$H_{i,t}=
(N_L^{i,t}H_L^{i,t}+N_R^{i,t}H_R^{i,t})/N$, where
$N=N_L^{i,t}+N_R^{i,t}$ and
$H_{\ell}^{i,t}
=-\sum_{j=0}^{1}p_{j,\ell}^{i,t}\log p_{j,\ell}^{i,t}$ for
$\ell\in\{L,R\}$.

The DFT score is
\begin{equation}
H_i^{\mathrm{DFT}}
=
\min_{t\in\mathcal{T}_i}
H_{i,t},
\label{eq_dft}
\end{equation}
where $\mathcal{T}_i$ denotes the candidate-threshold set.
A smaller $H_i^{\mathrm{DFT}}$ indicates stronger discrimination.
The best $D_s$ features form
$\tilde{\bm{f}}\in\mathbb{R}^{D_s}$.

All DFT scores and selected feature indices are determined using the
training set only and are fixed during inference, thereby avoiding
test-label leakage.

\subsection{XGBoost Decision Learning}

XGBoost performs the final nonlinear classification using
$\tilde{\bm{f}}$~\cite{chen2016xgboost}. Let $B$ denote the number
of decision trees and let $h_b(\tilde{\bm{f}})$ denote the output of
tree $b$. The ensemble score is
\begin{equation}
\xi(\tilde{\bm{f}})
=
\sum_{b=1}^{B}
h_b(\tilde{\bm{f}}).
\label{eq_xgb}
\end{equation}
The spoofing probability is
$\hat{p}=[1+\exp(-\xi(\tilde{\bm{f}}))]^{-1}$, and the final decision
is $\hat{c}=1$ if $\hat{p}\geq0.5$ and $\hat{c}=0$ otherwise.

SAAB, DFT, and XGBoost therefore provide complementary roles:
spatial-temporal representation, discriminative feature selection, and
nonlinear decision learning, respectively.


\section{Simulation Results}

\subsection{Simulation Setup}

Table~\ref{tab_parameter} summarizes the simulation and learning
parameters. The evaluation uses a balanced dataset of 20,000
physically feasible trajectories, including six VU motion patterns
with straight driving and lane changes. Equal numbers of legitimate
and spoofed trajectories are generated, and 20\% of the samples are
reserved for testing.

\begin{table}[t]
\caption{Simulation and Learning Parameters}
\centering
\setlength{\tabcolsep}{7pt}
\begin{tabular}{lc}
\toprule
\textbf{Parameter} & \textbf{Value} \\
\midrule
Carrier frequency & 28 GHz \\
Transmit antennas, $N_t$ & 32 \\
Receive antennas, $N_r$ & 32 \\
Transmit power, $P$ & 30 dBm \\
VU speed & 10 m/s \\
Sensing slots, $K$ & 150 \\
Slot duration, $T$ & 10 ms \\
Total trajectories & 20,000 \\
Legitimate/spoofed ratio & 1:1 \\
Test ratio & 20\% \\
RIS elements, $M$ & 8, 16, 32 \\
RIS locations & $(5,15),(5,20),(5,25)$ m \\
Noise power & $-100$ to $-60$ dBm \\
XGBoost trees, $B$ & 100 \\
Maximum tree depth & 3 \\
Learning rate & 0.7 \\
Subsampling ratio & 0.8 \\
Column sampling ratio & 0.8 \\
\bottomrule
\end{tabular}
\label{tab_parameter}
\end{table}

The spoofing data are generated using the physically grounded attack
framework in~\cite{shui2025spoofing}. At each sensing slot, the
attacker selects a feasible Doppler offset from $\mathcal{A}_k$ while
maintaining spatial-temporal consistency of the resulting fake
trajectory. In~\cite{shui2025spoofing}, this sequential attack-generation problem
is formulated as an MDP and solved using proximal policy optimization
(PPO) with action masking~\cite{schulman2017ppo}. The resulting
attack generator is used here to construct the corresponding
signal-level spoofed echoes.

The proposed GL detector is compared with three baselines:

\emph{1) LSTM:}
The LSTM detector~\cite{hochreiter1997lstm} serves as a signal-level
recurrent baseline. It uses 64-dimensional real/imaginary echo
features, 128 hidden units, two recurrent layers, 0.3 dropout, a
learning rate of $10^{-3}$, and a batch size of 64.

\emph{2) BiLSTM:}
The BiLSTM detector~\cite{schuster1997bidirectional} processes the
echo sequence bidirectionally and uses the same hyperparameters as
LSTM for a consistent comparison.

\emph{3) STL:}
The STL detector~\cite{shui2025spoofing} serves as the
trajectory-level baseline and detects attacks from reconstructed
spatiotemporal trajectories rather than directly from received echoes.


\subsection{Detection Performance}

\begin{figure}[t]
\centering

\begin{minipage}[t]{0.48\columnwidth}
\centering
\includegraphics[width=\linewidth]
{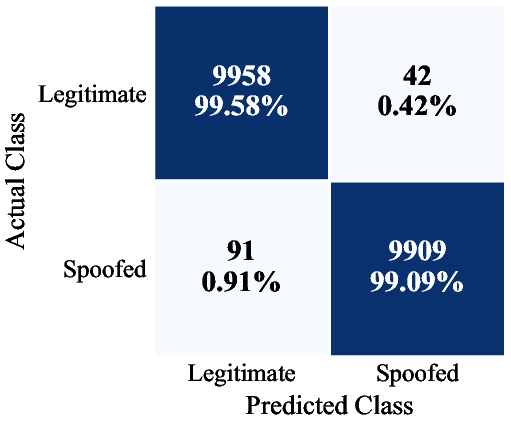}

\vspace{-0.05in}
\small (a)
\end{minipage}
\hfill
\begin{minipage}[t]{0.48\columnwidth}
\centering
\includegraphics[width=\linewidth]
{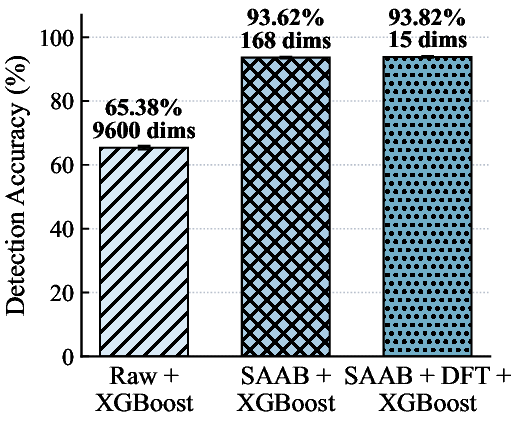}

\vspace{-0.05in}
\small (b)
\end{minipage}

\caption{Performance evaluation of the proposed GL detector:
(a) detection performance under $M=32$ and $-80$ dBm receiver
noise power; and
(b) ablation study of the GL architecture under $M=16$ and
$-80$ dBm receiver noise power.}
\label{fig_main}
\vspace{-0.08in}
\end{figure}

Fig.~\ref{fig_main}(a) shows the detection performance of the proposed
GL detector under $M=32$ and receiver noise power $-80$ dBm.
Aggregating the results over five random seeds, the proposed detector
achieves 99.335\% accuracy and approximately 99.58\% precision.
Among 20,000 test samples, 9,958 of 10,000 legitimate samples and
9,909 of 10,000 spoofed samples are correctly classified.

Let $\mathrm{TP}$, $\mathrm{TN}$, $\mathrm{FP}$, and $\mathrm{FN}$
denote the true-positive, true-negative, false-positive, and
false-negative sample counts, respectively. The detection,
missed-detection, and false-alarm probabilities are
$P_D=\mathrm{TP}/(\mathrm{TP}+\mathrm{FN})$,
$P_{\mathrm{MD}}=\mathrm{FN}/(\mathrm{TP}+\mathrm{FN})$, and
$P_{\mathrm{FA}}=\mathrm{FP}/(\mathrm{FP}+\mathrm{TN})$,
respectively. The proposed detector achieves
$P_D=99.09\%$, $P_{\mathrm{MD}}=0.91\%$, and
$P_{\mathrm{FA}}=0.42\%$.

Avoiding missed attacks is particularly important in vehicular
sensing security. A false alarm may trigger an unnecessary defensive
action, whereas an undetected spoofing attack can propagate incorrect
range, velocity, or trajectory information to downstream decision and
control functions.


\subsection{Ablation Study}

Fig.~\ref{fig_main}(b) presents the ablation results for $M=16$,
$-80$ dBm receiver noise power, and RIS position $(5,15)$ m.
Direct XGBoost on the original 9,600-dimensional representation
achieves 64.58\% accuracy. After SAAB, the feature dimension is
reduced to 168, corresponding to a 98.25\% reduction, while the
accuracy increases substantially to 93.62\%.

DFT further reduces the representation from 168 to only 15 features,
an additional reduction of approximately 91.1\%, while slightly
improving the accuracy to 93.82\%. Overall, the complete pipeline
compresses the original representation from 9,600 to 15 features,
exceeding 99.8\% dimensionality reduction, while improving the
detection accuracy by more than 29 percentage points over direct
XGBoost.


\subsection{Robustness to Receiver Noise}

\begin{figure*}[t]
\centering

\begin{minipage}[t]{0.32\textwidth}
\centering
\includegraphics[width=\linewidth]
{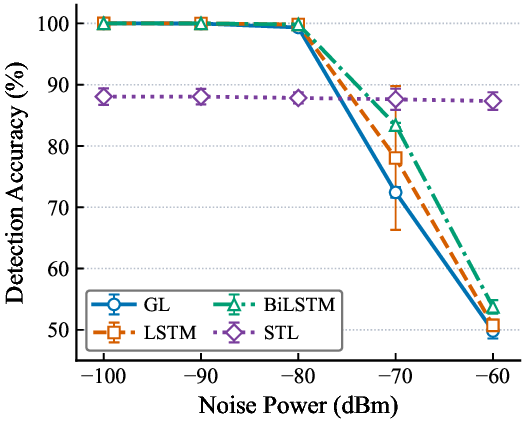}

\vspace{-0.05in}
\small (a)
\end{minipage}
\hfill
\begin{minipage}[t]{0.32\textwidth}
\centering
\includegraphics[width=\linewidth]
{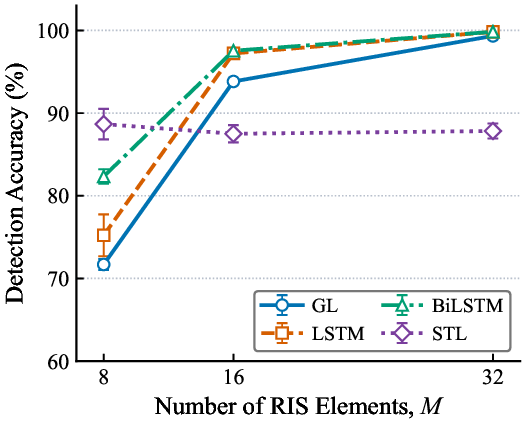}

\vspace{-0.05in}
\small (b)
\end{minipage}
\hfill
\begin{minipage}[t]{0.32\textwidth}
\centering
\includegraphics[width=\linewidth]
{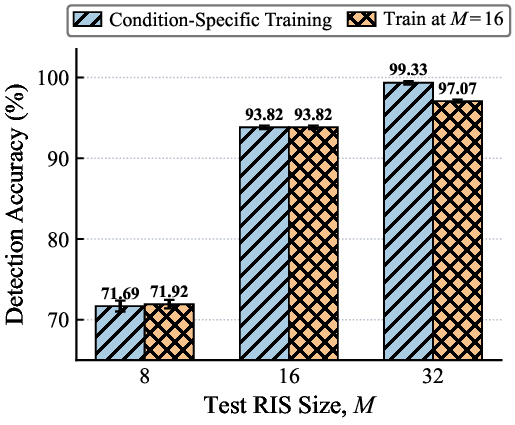}

\vspace{-0.05in}
\small (c)
\end{minipage}

\caption{Detection performance under different operating conditions:
(a) accuracy versus receiver noise power with $M=32$;
(b) accuracy versus the number of malicious RIS elements at
$-80$ dBm receiver noise power; and
(c) cross-condition generalization when the GL detector is trained
with $M=16$ and tested with unseen RIS sizes.}
\label{fig_generalization}
\vspace{-0.08in}
\end{figure*}

Fig.~\ref{fig_generalization}(a) shows the detection accuracy versus
receiver noise power from $-100$ to $-60$ dBm with $M=32$.
Receiver noise reduces the physical observability of the malicious
RIS echo and therefore directly affects signal-level detection.

At low-to-moderate noise powers, GL, LSTM, and BiLSTM achieve high
accuracy and outperform the trajectory-level STL detector because the
spoofing signature remains directly observable in the matched-filtered
echo. As receiver noise increases, the separability between
legitimate and spoofed waveform structures decreases, progressively
degrading the signal-level methods.

STL exhibits a different robustness profile because it evaluates
higher-level spatial-temporal consistency after trajectory
reconstruction. It can therefore become relatively more robust when
the physical-layer spoofing signature is severely corrupted. These
results show that signal-level and trajectory-level defenses are
complementary: signal-level learning enables earlier detection when
the malicious echo is observable, whereas trajectory-level reasoning
can provide additional robustness under severe waveform degradation.


\subsection{Impact of Malicious RIS Size}

Fig.~\ref{fig_generalization}(b) compares the detection accuracy for
$M\in\{8,16,32\}$ at a fixed receiver noise power of $-80$ dBm.
The proposed GL detector improves from 71.69\% at $M=8$ to
93.82\% at $M=16$ and 99.33\% at $M=32$. LSTM and BiLSTM
exhibit a similar increasing trend and also approach nearly perfect
detection at $M=32$, whereas STL is less sensitive to the increase
in RIS size.

This trend is consistent with the approximately $M^2$ coherent
reflection gain of the spoofing model. A larger malicious RIS
produces a stronger artificial echo relative to both the legitimate
VU echo and receiver noise, making the attack more effective against
the sensing estimator while simultaneously leaving a stronger
signal-level signature. Thus, increasing the number of reflecting
elements creates an inherent security trade-off for the attacker:
greater spoofing capability also increases physical observability.


\subsection{Cross-Condition Generalization}

Fig.~\ref{fig_generalization}(c) evaluates cross-condition
generalization across unseen RIS sizes. The GL detector is trained
only with $M=16$ and directly tested at $M=8$ and $M=32$
without retraining. The cross-condition detector achieves 71.82\%
accuracy at $M=8$ and 97.07\% at $M=32$, compared with
71.69\% and 99.33\%, respectively, under condition-specific training.

The two models therefore exhibit nearly identical performance at
$M=8$, with only a 0.13-percentage-point difference, whereas the
cross-condition model incurs a 2.26-percentage-point loss at
$M=32$. These results indicate useful zero-retraining transfer across
different RIS sizes, while the larger gap at $M=32$ reveals a
measurable distribution shift and motivates future multi-condition
training.


\subsection{Computational and Energy Efficiency}

The primary advantage of the proposed GL detector is its low
inference cost compared with recurrent neural networks.
Table~\ref{tab_complexity} reports both arithmetic complexity and
measured energy consumption.
All energy measurements were obtained on an AMD Ryzen 7 5700X CPU
and an NVIDIA GeForce RTX 3070 Ti GPU using CodeCarbon~\cite{courty2024codecarbon}.

\begin{table}[t]
\caption{Inference Complexity and Energy Consumption}
\centering
\setlength{\tabcolsep}{2.8pt}
\begin{tabular}{lccc}
\toprule
\textbf{Method} &
\textbf{FLOPs} &
\textbf{Energy (kWh)} &
\textbf{CO$_2$ (kg)}\\
\midrule
GL
& 0.1119 M
& $1.13{\times}10^{-4}$
& $7.28{\times}10^{-5}$\\
LSTM
& 68.8 M
& $1.24{\times}10^{-3}$
& $8.01{\times}10^{-4}$\\
BiLSTM
& 196.8 M
& $5.99{\times}10^{-4}$
& $3.85{\times}10^{-4}$\\
\bottomrule
\end{tabular}
\label{tab_complexity}
\end{table}

GL requires only 0.1119 million FLOPs, corresponding to approximately
615--1759 times fewer FLOPs than the considered recurrent baselines.
Using CodeCarbon~\cite{courty2024codecarbon}, GL consumes
$1.13\times10^{-4}$ kWh and produces
$7.28\times10^{-5}$ kg of CO$_2$, whereas LSTM and BiLSTM consume
$1.24\times10^{-3}$ kWh and $5.99\times10^{-4}$ kWh,
respectively.

The reduction in measured energy does not scale linearly with the
FLOP reduction because memory access, software overhead, processor
utilization, and runtime scheduling also contribute to system-level
energy consumption. Nevertheless, both arithmetic and measured energy
results consistently favor the GL architecture, which preserves
competitive detection accuracy with a substantially smaller
computational footprint.


\subsection{Discussion and Limitations}

Signal-level detection performance is primarily governed by the
physical observability of the spoofing echo: increasing $M$ strengthens
signal-level separability, whereas increasing receiver noise
suppresses the features that distinguish legitimate and spoofed
observations.

The present evaluation is limited to a single VU and a single
malicious RIS, with channel, mobility, and attack configurations
inherited from~\cite{shui2025spoofing}. Performance under multiple
targets, multiple malicious RISs, non-ideal hardware, channel-model
mismatch, and heterogeneous deployments remains to be investigated.


\section{Conclusion}

This paper proposed a lightweight signal-level GL framework for
detecting malicious RIS spoofing directly from matched-filtered RSU
echoes. The SAAB--DFT--XGBoost detector achieves 99.335\% accuracy
under $M=32$ and $-80$ dBm receiver noise power, generalizes across unseen
RIS sizes, and substantially reduces computational and measured
energy costs relative to recurrent baselines.

The results show that signal-level detectability depends on the
physical observability of the spoofing echo: larger RISs improve
separability, whereas receiver noise degrades it. Future work will
consider multiple VUs and malicious RISs, channel-model mismatch,
hardware impairments, and joint signal- and trajectory-level
detection.



\begin{thebibliography}{99}

\bibitem{itur2160}
ITU-R, ``Framework and overall objectives of the future development
of IMT for 2030 and beyond,'' International Telecommunication Union,
Recommendation ITU-R M.2160-0, Nov. 2023. [Online]. Available:
https://www.itu.int/rec/R-REC-M.2160-0-202311-I/en

\bibitem{isacsurvey2025}
X. Luo, Q. Lin, R. Zhang, H.-H. Chen, X. Wang, and M. Huang,
``ISAC---a survey on its layered architecture, technologies,
standardizations, prototypes, and testbeds,''
\emph{IEEE Commun. Surveys Tuts.}, vol. 28, pp. 485--526, 2026.

\bibitem{du2025vehicular}
Z. Du, F. Liu, Y. Li, W. Yuan, Y. Cui, Z. Zhang, C. Masouros, and B. Ai,
``Toward ISAC-empowered vehicular networks: Framework, advances,
and opportunities,''
\emph{IEEE Wireless Commun.}, vol. 32, no. 2, pp. 222--229,
Apr. 2025.

\bibitem{su2021secure}
N. Su, F. Liu, and C. Masouros,
``Secure radar-communication systems with malicious targets:
Integrating radar, communications and jamming functionalities,''
\emph{IEEE Trans. Wireless Commun.}, vol. 20, no. 1, pp. 83--95,
Jan. 2021.

\bibitem{su2022secure}
N. Su, F. Liu, Z. Wei, Y.-F. Liu, and C. Masouros,
``Secure dual-functional radar-communication transmission:
Exploiting interference for resilience against target eavesdropping,''
\emph{IEEE Trans. Wireless Commun.}, vol. 21, no. 9,
pp. 7238--7252, Sep. 2022.

\bibitem{cao2025sensing}
Y. Cao, L. Duan, and R. Zhang,
``Sensing for secure communication in ISAC: Protocol design and
beamforming optimization,''
\emph{IEEE Trans. Wireless Commun.}, vol. 24, no. 2,
pp. 1207--1220, Feb. 2025.

\bibitem{chrysanidis2024replay}
G. Chrysanidis, Y. Liu, and A. Argyriou,
``A replay attack against ISAC based on OFDM,''
\emph{IEEE Access}, vol. 12, pp. 20998--21003, 2024.

\bibitem{xu2025stealth}
F. Xu, W. Lai, and K. Shen,
``Intelligent surface assisted radar stealth against unauthorized ISAC,''
\emph{IEEE Wireless Commun. Lett.}, vol. 14, no. 4,
pp. 1149--1153, Apr. 2025.

\bibitem{hassouna2026riscs}
S. Hassouna, M. Zubair, A. Raza, M. Ahmed, J. Kazim,
M. A. Imran, and Q. H. Abbasi,
``RISCS: RIS-aided integrated sensing and communications---a
cyber-physical security perspective,''
\emph{IEEE Trans. Consum. Electron.}, vol. 72, no. 3,
pp. 7579--7601, Aug. 2026.

\bibitem{shui2025safety}
T. Shui, W. Saad, and M. Chen,
``Sensing safety analysis for vehicular networks with integrated
sensing and communication (ISAC),''
in \emph{Proc. IEEE Global Commun. Conf. (GLOBECOM)},
Taipei, Taiwan, 2025, pp. 775--780.

\bibitem{shui2025spoofing}
T. Shui, P.-H. Chou, W. Saad, and M. Chen,
``Analysis and detection of RIS-based spoofing in integrated sensing
and communication (ISAC),''
\emph{arXiv preprint arXiv:2508.18100}, 2025.

\bibitem{hochreiter1997lstm}
S. Hochreiter and J. Schmidhuber,
``Long short-term memory,''
\emph{Neural Comput.}, vol. 9, no. 8, pp. 1735--1780,
Nov. 1997.

\bibitem{kuo2019interpretable}
C.-C. J. Kuo, M. Zhang, S. Li, J. Duan, and Y. Chen,
``Interpretable convolutional neural networks via feedforward design,''
\emph{J. Vis. Commun. Image Represent.}, vol. 60, pp. 346--359,
Apr. 2019.

\bibitem{yang2022supervised}
Y. Yang, W. Wang, H. Fu, and C.-C. J. Kuo,
``On supervised feature selection from high dimensional feature spaces,''
\emph{APSIPA Trans. Signal Inf. Process.}, vol. 11, p. e31, 2022.

\bibitem{kuo2022green}
C.-C. J. Kuo and A. M. Madni,
``Green learning: Introduction, examples and outlook,''
\emph{J. Vis. Commun. Image Represent.}, vol. 90, p. 103685,
Feb. 2023.

\bibitem{chen2016xgboost}
T. Chen and C. Guestrin,
``XGBoost: A scalable tree boosting system,''
in \emph{Proc. 22nd ACM SIGKDD Int. Conf. Knowl. Discovery Data Mining},
San Francisco, CA, USA, Aug. 2016, pp. 785--794.

\bibitem{schuster1997bidirectional}
M. Schuster and K. K. Paliwal,
``Bidirectional recurrent neural networks,''
\emph{IEEE Trans. Signal Process.}, vol. 45, no. 11,
pp. 2673--2681, Nov. 1997.

\bibitem{schulman2017ppo}
J. Schulman, F. Wolski, P. Dhariwal, A. Radford, and O. Klimov,
``Proximal policy optimization algorithms,''
\emph{arXiv preprint arXiv:1707.06347}, 2017.

\bibitem{courty2024codecarbon}
B. Courty, V. Schmidt, \emph{et al.},
``mlco2/codecarbon: v2.4.1,''
Zenodo, Software, Version 2.4.1, May 2024.

\end{thebibliography}
\end{document}